\documentclass[aps,prx,twocolumn]{revtex4-1}  	% use "amsart" instead of "article" for AMSLaTeX format
\usepackage{times}
\usepackage{graphicx}				% Use pdf, png, jpg, or eps\UTF{00C2}§ with pdflatex; use eps in DVI mode
\usepackage{amssymb,amsmath}
\usepackage{physics}
\begin{document}
\title{Discrete-time quantum walk on complex networks for community detection}
\author{Kanae Mukai}
\affiliation{
Department of Physics, University of Tokyo,
5-1-5 Kashiwanoha, Kashiwa, Chiba 277-8574, Japan}
%\vspace{1.0\baselineskip}\\
\author{Naomichi Hatano}
\affiliation{Institute of Industrial Science, University of Tokyo,
5-1-5 Kashiwanoha, Kashiwa, Chiba 277-8574, Japan}
%e-mail: \texttt{hatano@iis.u-tokyo.ac.jp}
%\date{\today}							% Activate to display a given date or no date
%\thispagestyle{empty}
%\subsection{}
%\setcounter{page}{0}
%\newpage

\begin{abstract}
Many systems such as social networks and biological networks take the form of complex networks, which have the community structure. The community detection in complex networks is of great interest for many researchers in statistical physics and mathematical physics. There have been studies on community detection that use the classical random walk. The present study utilizes the discrete-time quantum walk instead. The quantum walk plays an important role in various fields, especially in the research of quantum computers, and attracts much attention from mathematical physics too. The discrete-time quantum walk has two properties: it linearly spreads on a flat space;
%, which is quadratically faster than the classical random walk; 
it localizes in some cases because of quantum coherence. 
We demonstrate that these properties of the quantum walk are useful for community detection on complex networks. We define the discrete-time quantum walk on complex networks and utilize it for community detection. We numerically show that the quantum walk with the Fourier coin is localized in a community to which the initial node belongs. 
%The infinite-time average of the normalized transition probability between two nodes, calculated from the eigenvectors, reveals the community structure, indicating that the eigenvectors contain information about localization to communities. 
%We also find that the infinite-time average reveals the community structure better if the eigenvalues of the time-evolution unitary matrix are non-degenerate, and hence the Fourier walk is more suitable for community detection than the Grover walk. 
Meanwhile, the quantum walk with the Grover coin tends to be localized around the initial node, not over a community. The probability of the classical random walk on the same network converges to the uniform distribution with a relaxation time generally unknown \textit{a priori}. We thus claim that the time average of the probability of the Fourier-coin quantum walk on complex networks reveals the community structure more explicitly than that of the Grover-coin quantum walk and a snapshot of the classical random walk. We first demonstrate our method of community detection for a prototypical three-community network, producing the correct grouping. We then apply our method to two real-world networks, namely Zachary's karate club and the US Airport network,.
We successfully reveals the community structure, the two communities of the instructor and the administrator in the former and major airline companies in the latter.
%; namely Zachary's karate-club network and the neural network of C. elegans. We 
%and show that our results are roughly consistent with those by other methods.
%For Zachary's karate-club network, we show that our method reveals its community structure correctly. For the neural network of C. elegans, we show that our result is roughly consistent with that by the modularity-based method.
\end{abstract}

\maketitle

\section{Introduction}
\label{sec1}
\subsection{Quantum walk}
\label{sec1A}
The quantum walk has been studied in various areas of physics. The quantum walk is divided into two types: the discrete-time quantum walk~\cite{quantumwalkdefinition} and the continuous-time quantum walk~\cite{CTQW}. The time evolution of the latter is expressed by a Hamiltonian obeying the Schr\"odinger equation. In the present paper, we focus on the former. 

The discrete-time quantum walk is a quantum counterpart of the discrete-time classical random walk. In the classical random walk \textit{e.g.~}in one dimension, a particle hops to the left or right stochastically, generating a probability distribution, whereas the quantum walk is described instead in terms of the probability amplitude of quantum superposition of the left-mover and the right-mover~\cite{quantumwalkdefinition}. 

The quantum walk generally has the following two properties: it linearly spreads on a flat space and localizes in particular spots~\cite{quantumwalkbasic}. To be more specific, however, quantum walks with different inner states and different coin operators behave differently. The probability distribution of the three-state quantum walk in one dimension, for example, has three peaks, one that moves linearly to the left, one that moves linearly to the right, and the one that localizes at the initial node~\cite{threestate}. The one of the two-state quantum walk in one dimension, on the other hand, has only two peaks that spread linearly to the left and right, without any peak that localizes~\cite{twostate}. In the present thesis, we focus on the two-state walk, using the Fourier coin and the Grover coin~\cite{grover, FQWGQW}. The walks with these coin operators are called the Fourier walk and the Grover walk, respectively. We will demonstrate that the two walks behave differently.

The quantum walk has been applied to quantum computers, search problems and so on~\cite{quantumcomputer, shiftronbun, implementing}. Many researchers consider that the quantum-mechanical computers may solve problems more efficiently than the classical computers. The quantum walk has been already implemented in the laboratory~\cite{laboratory}. 

There have been several studies on the quantum walk on networks, mostly on regular ones~\cite{shiftronbun, directed}. The shift operator and the coin operator have been defined in conformity to the structure of networks. The quantum walk on networks occupies an important role on search problems. In general, it takes classical algorithms $O(N)$ steps to identify the target record from an unsorted database of $N$ records, while it takes quantum mechanical systems only $O(\sqrt{N})$ steps~\cite{quantumcomputer}.

\subsection{Complex networks}
\label{sec1B}
Many systems including social networks and biological networks have been found to have distinctive features drastically different from random graphs, and hence are collectively called complex networks~\cite{senseikara,review-estrada,review-chen,review-latora}. 
Representative examples include acquaintance networks~\cite{zachary}, the World Wide Web~\cite{WorldWideWeb}, corporate transaction networks~\cite{takayasu}, neural networks~\cite{neuralnetwork}, food webs~\cite{foodweb} and metabolic networks~\cite{metabolicnetwork}. 

The distinctive features of the complex networks often quoted include the scale-free property and the small-world effect, although there are so-called complex networks that do not have these features.
The former feature means that the histogram of the degrees of the nodes (the number of the links attached to a node) follows a power-law behavior~\cite{powerlaw}; in other words, there are a large number of nodes with low degrees and a small number of nodes with high degrees in a self-similar way.
The latter feature means that the average distance between randomly chosen pair of nodes in a complex network is surprisingly shorter than that in a random network~\cite{smallworld}.

\begin{figure}
\centering
\includegraphics[width=\columnwidth]{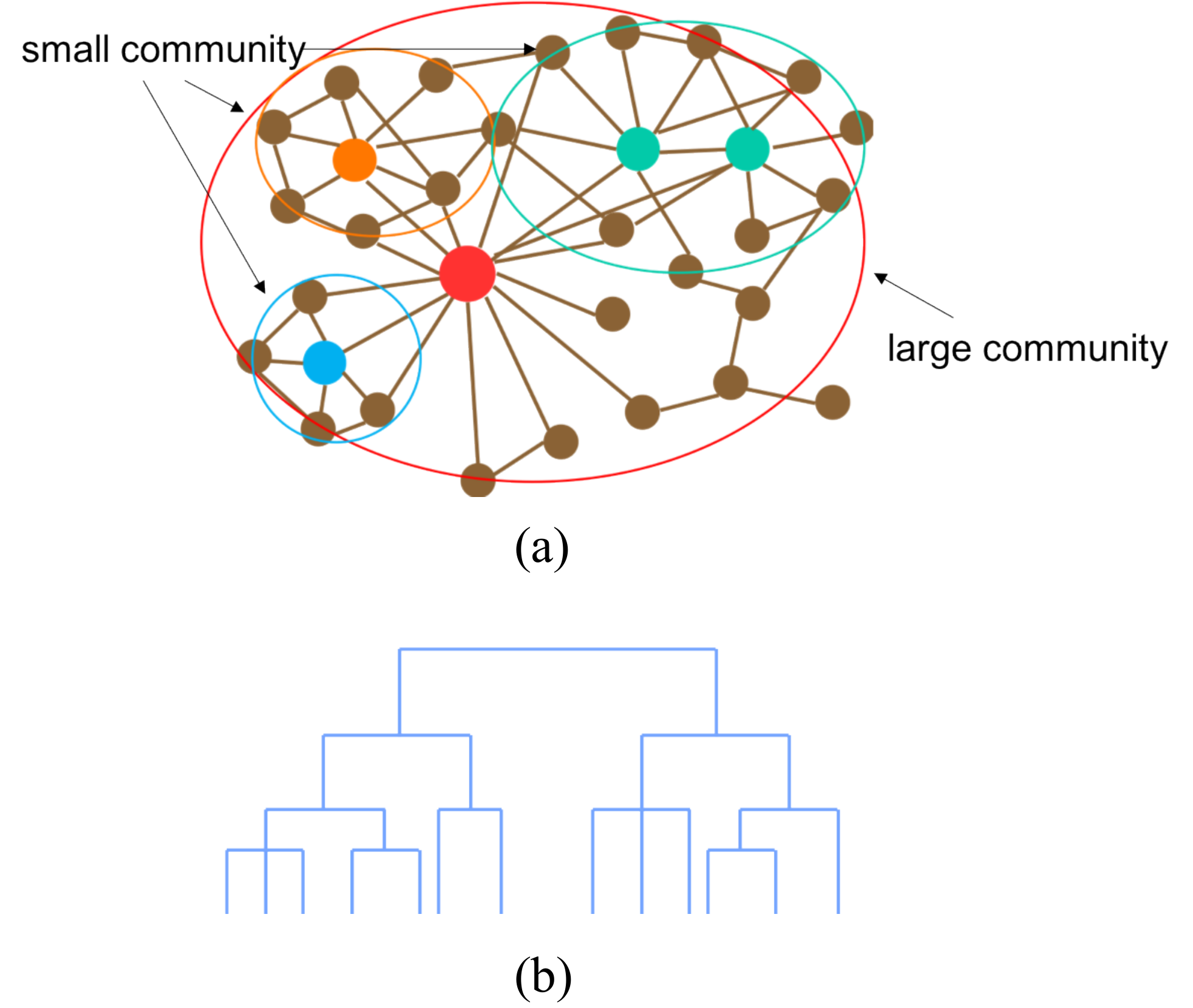}
\caption{(a) A schematic illustration of the communities of a complex network arranged in a hierarchy. (b) An example of a dendrogram.}
\label{dendrogramcommunity}
\end{figure}

These features may indicate that many complex networks have a hierarchical structure; see Fig.~\ref{dendrogramcommunity}, for example.
When we depict the structure as a tree, which is called a dendrogram in the social sciences~\cite{communitydefinition} (see Fig.~\ref{dendrogramcommunity}~(b), for example), the leaves correspond to the nodes and the branches to the links.
Nodes in higher levels of the dendrogram can have more links to nodes in lower levels in a self-similar way.
A node in one branch of the dendrogram to another node in a different branch can be connected by a short path through nodes in higher levels.

In a hierarchical complex network, we should be able to find communities in various levels.
The community is a subset of nodes within the network such that connections among the nodes of the community are denser than those among the other nodes~\cite{communitydefinition}.
As the hierarchy in Fig.~\ref{dendrogramcommunity} suggests, a node at a high level of the dendrogram is likely to be at the center of each community typically with many links, which we call a hub.
It is therefore of great importance for detecting the features of complex networks to identify communities.

There are several algorithms for community detection~\cite{senseikara, communitydefinition, mathematica, Newman, usingrandomwalk}. The conventional method is the hierarchical clustering~\cite{senseikara, communitydefinition, mathematica, book}. In this method, one calculates a weight $W_{i,j}$ for every pair of nodes in the network. The weight shows how closely connected the nodes are. Starting from the nodes with no links between them, one adds links between pairs in the order of their weights. The nodes are classified into communities, and the communities are grouped into larger communities. Many different weights have been proposed in this algorithm. The weight considering the paths longer than the shortest ones was taken into account in Ref.~\cite{naomichi}. Another method is called the divisive algorithm~\cite{senseikara}. Starting from the whole network, one cuts the links. The network is divided into smaller subnetworks, which are identified as communities. Another research presents an algorithm with a modularity~\cite{mathematica, Newman, Modularity, randomcommunityorder}. The modularity is a property of a network and a division of the network into communities. If there are many links within the communities and a few links between the communities, the division is good. 

There have been several studies on community detection that used the discrete-time classical random walks~\cite{usingrandomwalk, randomcommunity}. These approaches are based on the consideration that random walks on the networks tend to get trapped within the communities~\cite{usingrandomwalk}. One computes the frequency in which each node is visited by a random walker, and explores the possible partitions by using deterministic algorithms~\cite{randomcommunity}. 

We here utilize the discrete-time $\it{quantum}$ walk instead for community detection. The infinite-time average of the transition probability, normalized by the number of links, of the Fourier-coin quantum walk on a complex network shows localization in a community, and thereby reveals the community structure. The Grover-coin quantum walk, in contrast, tends to be localized around the initial node, presumably due to the localized eigenstates of the time-evolution unitary with degenerate $\pm 1$ eigenvalues. For the classical random walk on the same network, the probability converges to a flat distribution as time passes. Although the community structure partially emerges before the convergence, it is generally \textit{a priori} unknown which time step of the walk is best for the community detection. We thus claim that the Fourier-coin quantum walk on complex networks reveals the community structure more explicitly than the Grover-coin quantum walk and the classical random walk. 

\section{Quantum walk on complex networks}
\label{sec2}
We first describe our definition of the quantum walk on complex networks. It requires a node-dependent coin operator because each node has generally a different number of links.

\begin{figure}
\centering
\includegraphics[width=0.4\columnwidth]{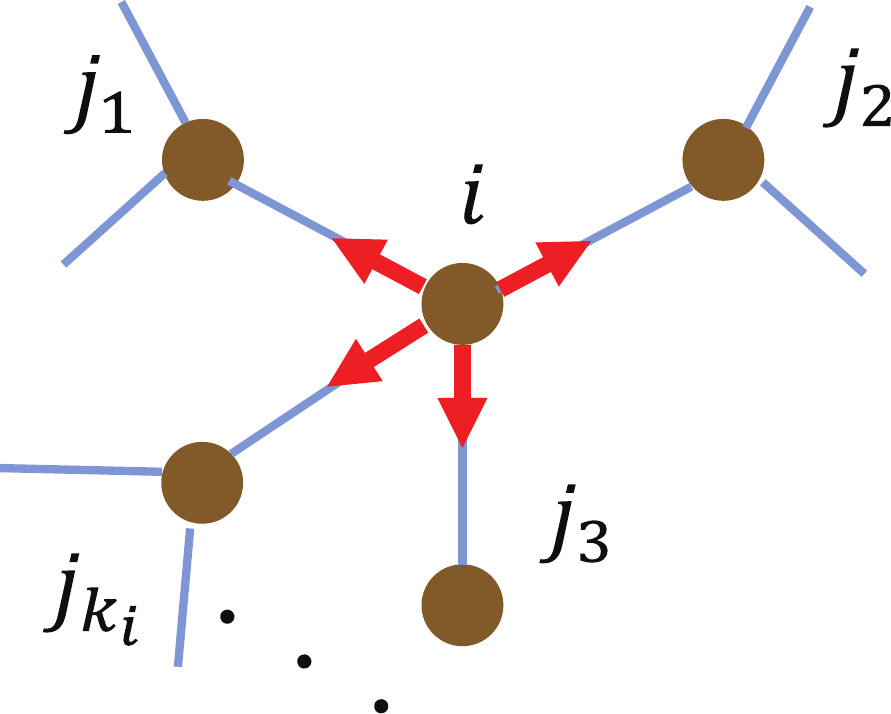}
\caption{The definition of a quantum state on complex networks.}
\label{quantumstate}
\end{figure}

We define the quantum state on a complex network (see Fig.~\ref{quantumstate}, for example)
in the form
\begin{align}
\ket{\psi(t)}&=\sum_{i=1}^{N}\sum_{j=1}^{k_i}\psi_{i,j}(t)\ket{i\to j},
\end{align}
where $N$ is the total number of nodes, the state $\ket{i\to j}$ resides on the node $i$ and is about to hop to the adjacent node $j$ on a link connecting $i$ and $j$, while $k_i$ is the number of links attached to the node $i$. The total Hilbert space $\mathcal H=\mathcal H_1\oplus\mathcal H_2\oplus\cdots\oplus\mathcal H_N$ consists of the Hilbert space of each node $\mathcal H_i$, which is spanned by $\left(\ket{i\to j_1},\ket{i\to j_2},\cdots,\ket{i\to j_{k_i}}\right)$. The dimensionality of the total Hilbert space is therefore given by
\begin{align}
D=\sum_{i=1}^{N}k_i,
\end{align} 
which is the total number of links under double counting.
We normalize the state $\ket{\psi(t)}$ as in
\begin{align}
\braket{\psi(t)}{\psi(t)}=\sum_{i=1}^{N}\sum_{j=1}^{k_i}\abs{\psi_{i,j}(t)}^2=1.
\end{align}
We can write the probability of the existence on a node $i$ at time $t$ as
\begin{align}
p(i;t)=\sum_{j=1}^{k_i}\abs{\psi_{i,j}(t)}^2.
\end{align}

The time evolution of the state $\ket{\psi(t)}$ is given by
\begin{align}
\label{timeevolution}
\ket{\psi(t)}&=U\ket{\psi(t-1)}\\
&=U^t\ket{\psi(0)},
\end{align}
where the unitary operator $U$ is the product of a shift operator $S$ and a coin operator $C$:
\begin{align}
U=SC.
\label{unitarymatrix}
\end{align}
We define the shift operator $S:\mathcal H\to\mathcal H$ by
\begin{align}
S\ket{i\to j}=\ket{j\to i}.
\label{shiftdefinition}
\end{align}
The choice of this shift operator may appear to be atypical compared to the one defined for the one-dimensional lattice, but it is necessary because of the existence of dangling bonds. When the node $j$ is at the end of a dangling bond as the bottom one $i\to j_{3}$ in Fig.~\ref{quantumstate}, Eq.~\eqref{shiftdefinition} is the only possible choice. Indeed, it has been used for searching a marked vertex on a specific graph called the Cayley tree~\cite{shiftronbun}.
We can also easily prove that the shift operator of Eq.~\eqref{shiftdefinition}, if defined on a one-dimensional lattice, can be mapped to the standard shift operator by introducing an extra factor to the coin operator; see Appendix~\ref{AppA}.

We define the coin operator $C$ by
\begin{align}
C&=C_1\oplus C_2\oplus\cdots\oplus C_N,
\end{align}
where we first set the coin operator of a node $i$, $C_i:\mathcal H_i\to\mathcal H_i$ as
\begin{widetext}
\begin{align}
&C^{\mathrm{F}}_{i}
\begin{pmatrix}
\ket{i\to j_1}\\
\ket{i\to j_2}\\
\ket{i\to j_3}\\
\vdots\\
\ket{i\to j_{k_i}}
\end{pmatrix}
%\nonumber\\
=\frac{1}{\sqrt{k_i}}\mqty(
1&1&1&\cdots&1\\
1&e^{i\theta/k_i}&e^{2i\theta/k_i}&\cdots&e^{(k_i-1)i\theta/k_i}\\
1&e^{2i\theta/k_i}&e^{4i\theta/k_i}&\cdots&e^{2(k_i-1)i\theta/k_i}\\
\vdots&\vdots&\vdots&\ddots&\vdots\\
1&e^{(k_i-1)i\theta/k_i}&e^{2(k_i-1)i\theta/k_i}&\cdots&e^{(k_i-1)(k_i-1)i\theta/k_i}
)
\begin{pmatrix}
\ket{i\to j_1}\\
\ket{i\to j_2}\\
\ket{i\to j_3}\\
\vdots\\
\ket{i\to j_{k_i}}
\end{pmatrix}
\end{align}
\end{widetext}
with $\theta=2\pi$. (Note that the numbering of the neighboring nodes $\{ j_1, j_2, \cdots, j_{k_i}\}$ is arbitrary but does affect the dynamics.) This specific operator is called the Fourier coin~\cite{FQWGQW} because it is a Fourier matrix. The Fourier-coin quantum walk (the Fourier walk) has been used on a particular kind of network~\cite{coinronbun}. 

Below we also consider the quantum walk with an alternative coin operator, namely the Grover coin~\cite{grover}, which is given by
\begin{widetext}
\begin{align}\label{GroverCoin}
&C^{\mathrm{G}}_{i}
\begin{pmatrix}
\ket{i\to j_1}\\
\ket{i\to j_2}\\
\ket{i\to j_3}\\
\vdots\\
\ket{i\to j_{k_i}}
\end{pmatrix}
%\nonumber\\
=\frac{1}{k_i}\mqty(
2-k_i&2&2&2&2\\
2&2-k_i&2&2&2\\
2&2&2-k_i&2&2\\
\vdots&\vdots&\vdots&\ddots&\vdots\\
2&2&2&\cdots&2-k_i
)
%=\frac{1}{1+\alpha}\mqty(
%-1&\alpha&\alpha&\cdots&\alpha\\
%\alpha&-1&\alpha&\cdots&\alpha\\
%\alpha&\alpha&-1&\cdots&\alpha\\
%\vdots&\vdots&\vdots&\ddots&\vdots\\
%\alpha&\alpha&\alpha&\cdots&-1
%)
\begin{pmatrix}
\ket{i\to j_1}\\
\ket{i\to j_2}\\
\ket{i\to j_3}\\
\vdots\\
\ket{i\to j_{k_i}}
\end{pmatrix}
\end{align}.
\end{widetext}
%with $\alpha$ being
%\begin{align}
%\alpha=\frac{2}{k_i-2}.
%\end{align}
This is called the Grover matrix, being related to Grover's search algorithm~\cite{algorithm}. There are many studies on the Grover-coin quantum walk (Grover walk). The periodicity of the Grover walk on some finite graphs has been clarified~\cite{finitegraph}. We will show that the Fourier coin works much better than the Grover coin for the purpose of community detection.

We prepare the initial state for the quantum walk as a state in which a specific state on a specific node $i_\textrm{start}$, $\ket{i_\textrm{start}\to j}$, has the element unity and the others have elements zero. In the next section~\ref{sec3}, we take the average over the adjacent nodes $j$ as will be seen in~\eqref{siteaverage} below.

\section{Community detection}
\label{sec3}
\subsection{Infinite-time average}
\label{sec3A}
We numerically show hereafter that the probability of the Fourier walk becomes higher in hubs as time passes whichever node we choose as the initial one $i_\textrm{start}$. We can thus detect hubs of complex networks, although the threshold to detect them is an open question. We will also show that the state of the Fourier walk on complex networks is localized in a community of the initial node, and thereby reveals the community structure. For the quantum walk on the one-dimensional finite lattice, the probability distribution after a long period of time has been proved to be stationary and uniform when the quantum walk behaves symmetrically~\cite{uniformconvergence}. For the quantum walk on complex networks, on the other hand, we here show that the infinite-time average of the normalized transition probability, calculated from the eigenvectors, shows localization.

Let us calculate the infinite-time average of the transition probability by expanding the unitary operator $U=SC$ in terms of its eigenstates:
\begin{align}
U=\sum_{\mu=1}^{D}\ket{\mu}e^{i\theta_{\mu}}\bra{\mu},
\end{align}
where $\ket{\mu}$ is the eigenvector and $e^{i\theta_{\mu}}$ is its eigenvalue with a real argument $\theta_\mu$. The transition probability that the quantum walk starting from a node $i$ reaches a node $l$ is given by
\begin{align}
\label{siteaverage}
p(i\to l;t)&=\frac{1}{k_i}\sum_{m=1}^{k_l}\sum_{j=1}^{k_i}\abs{\bra{l\to m}U^t\ket{i\to j}}^2,
\end{align}
where $\ket{i\to j}$ is the initial state and $\ket{l\to m}$ is the state at the step $t$. The factor $1/k_i$ is to average over the direction $j$ of the initial state. We also took the summation over the direction $m$ of the final state.

The infinite-time average of the transition probability is given by
\begin{align}
\overline{p(i\to l)}&=\lim_{T\to\infty}\frac{1}{T}\frac{1}{k_i}\sum_{t=0}^{T-1}\sum_{m=1}^{k_l}\sum_{j=1}^{k_i}\abs{\bra{l\to m}U^t\ket{i\to j}}^2
\\
&=\frac{1}{k_i}\sum_{\mu=1}^{D}\sum_{m=1}^{k_l}\sum_{j=1}^{k_i}\abs{\bra{l\to m}\ket{\mu}}^2\abs{\bra{\mu}\ket{i\to j}}^2,
\label{infinitetimeave}
\end{align}
where we assumed 
\begin{align}
\label{Kronecker}
\lim_{T\to\infty}\frac{1}{T}\sum_{t=0}^{T-1}e^{i(\theta_\mu-\theta_\nu)t}=\delta_{\mu \nu},
\end{align}
which is valid if the eigenvalues are non-degenerate and distributed almost randomly over the unit circle. In this case, the quantum walk on the network is a superposition of oscillation with various frequencies, and hence the infinite-time average makes sense. 

\begin{figure*}
\centering
\includegraphics[width=0.7\textwidth]{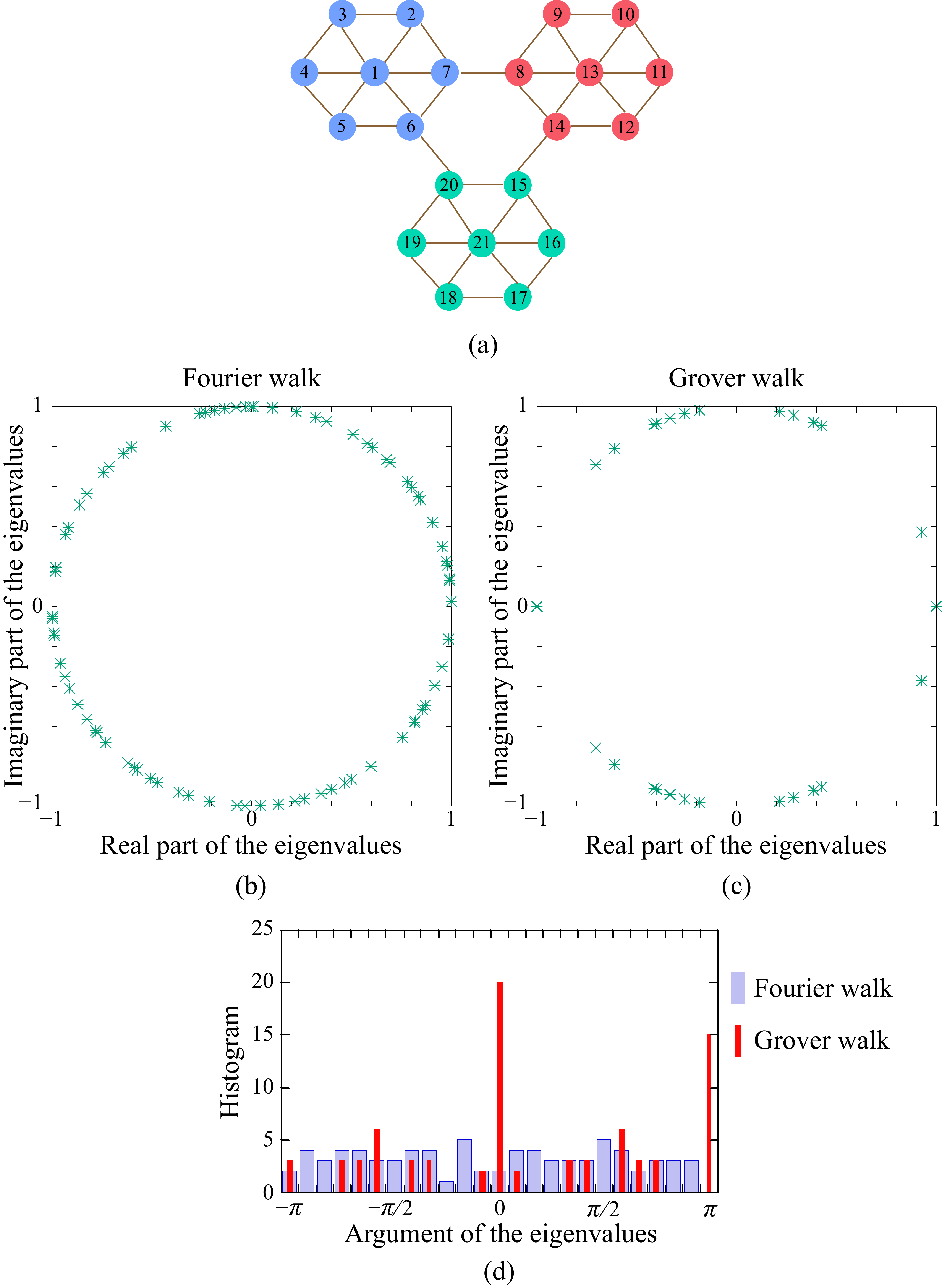}
\caption{
%(a) The definition of the quantum walk on complex networks. 
(a) A prototypical three-community network, for which $N=21$ and $D=78$. The hubs are the nodes 1, 13, and 21. (b) and (c) The complex eigenvalues of the time-evolution unitary matrix $U$ for (b) the Fourier walk and (c) the Grover walk on the three-community network in~(a). 
The horizontal axis shows the real part and the vertical axis shows the imaginary part of the eigenvalues. 
There are 78 eigenvalues plotted because $D=78$. 
Within the numerical double precision, the unitary matrix of the Fourier walk has no degeneracy in (b), whereas for the one of the Grover walk, the eigenvalues $\pm 1$ have the degeneracy 20 and 18, respectively, in (c). (d) The distribution of the argument of the eigenvalues for the Fourier walk (fat blue columns) and for the Grover walk (thin red columns).}
\label{three_matome}
\end{figure*}

In order to check the validity of the formulation, we show in Fig.~\ref{three_matome}~(b)--(c) the eigenvalue distributions of the time-evolution unitary matrix $U$ for the Fourier walk and the Grover walk on a prototypical three-community network given in Fig.~\ref{three_matome}~(a), for which $N=21$ and $D=78$.
In both cases, the 78 eigenvalues are distributed over a unit circle on the complex plane.
The eigenvalues of the Fourier walk are non-degenerate, while almost half of the eigenvalues of the Grover walk are degenerate either at $\pm1$. (Precisely, the degeneracies are 20 and 18 for the eigenvalues $\pm 1$, respectively.)
The histogram in Fig.~\ref{three_matome}~(d) shows more clearly that the eigenvalues of the Fourier walk are distributed much more evenly over the unit circle than the eigenvalues of the Grover walk.
We thus realize that the Fourier walk is more suitable for the formulation~\eqref{infinitetimeave} than the Grover walk.

It has been proven for the Grover walk that the eigenvectors of the eigenvalues degenerate to $\pm 1$ are localized on loops of graphs~\cite{Segawa14};
indeed the degree of the degeneracy is completely determined by the topology of the graph (see Appendix~\ref{AppB} for tutorial examples).
On regular graphs, this degeneracy would lead to a proof of the localization on the initial node after linear combination of the eigenvectors on the loops~\cite{Segawa14}.
We will numerically show below that the Grover walk on a graph is also localized around the initial node of the walk.

\begin{figure*}
\begin{center}
\includegraphics[width=0.8\textwidth]{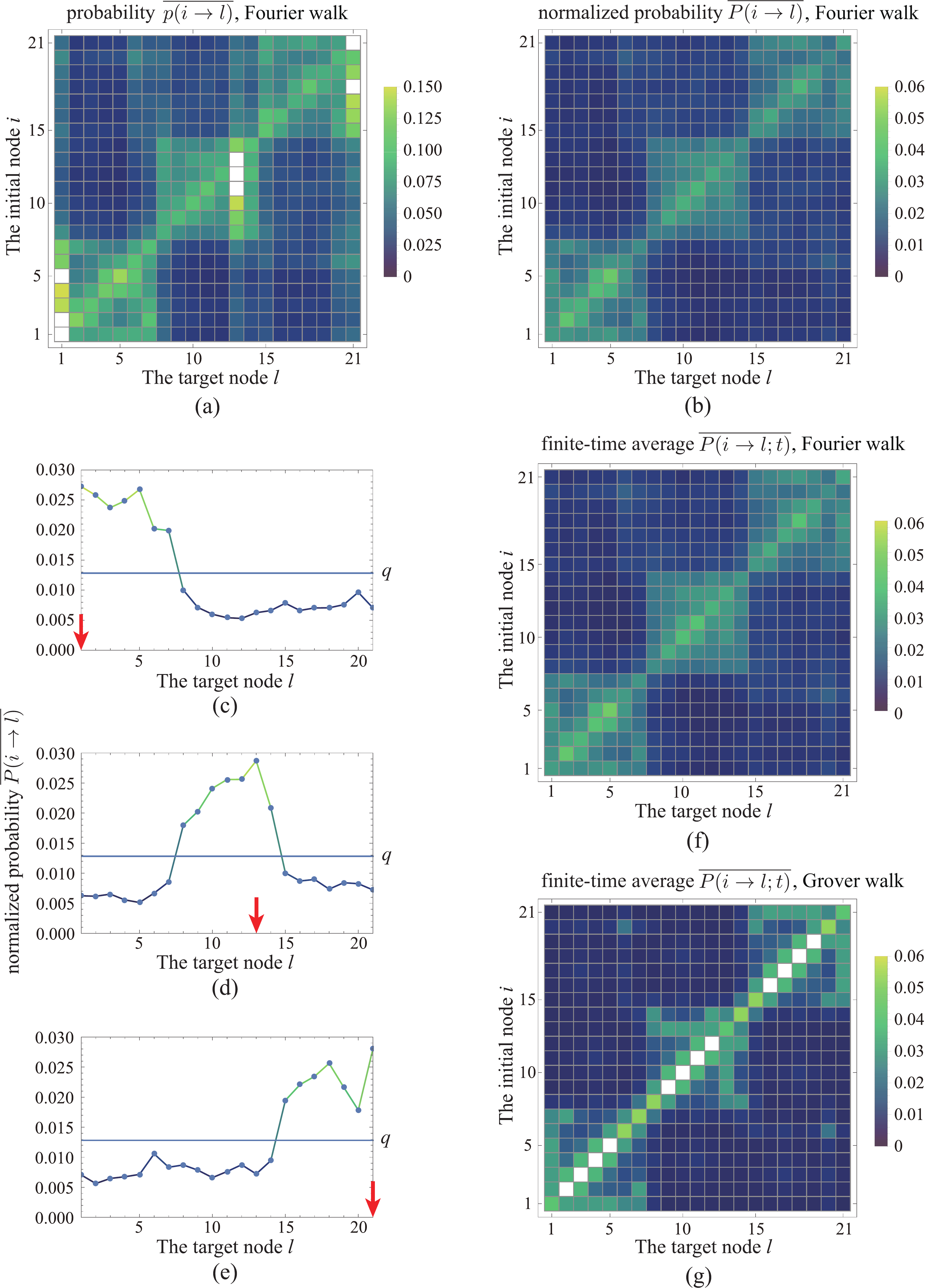}
\end{center}
\caption{(a) The infinite-time average~\eqref{infinitetimeave} of the probability $\overline{p(i\to l)}$ of the Fourier walk on the three-community network in Fig.~\ref{three_matome}~(a). (b) The infinite-time average of the \textit{normalized} probability $\overline{P(i\to l)}$ in Eq.~\eqref{normalizedprob} of the Fourier walk on the three-community network. In (b) and (c), the vertical axis shows the initial node $i$ and the horizontal axis shows the target node $l$, while each square indicates either value of $\overline{p(i\to l)}$ or $\overline{P(i\to l)}$. A white square indicates a value off scale in the higher direction. (c)--(e) The infinite-time average of the normalized probability $\overline{P(i\to l)}$ in Eq.~\eqref{normalizedprob} of the Fourier walk on the three-community network that starts from (c) the hub $i=1$, (d) the hub $i=13$, and (e) the hub $i=21$, each of which is indicated by a red arrow. The horizontal axis shows the target node $l$ and the vertical axis shows the normalized probability $\overline{P(i\to l)}$ with $i=1$, $13$, $21$. The horizontal line in the middle indicates the threshold $q=1/78\simeq 0.0128$. (f) The time average of the normalized probability $P(i\to l;t)$ over the first 100 steps of the Fourier walk on the three-community network. (g) The same but for the Grover walk.
%The time average of the normalized probability $P(i\to l;t)$ over 100 steps of the Grover walk on the three-community network.
}
\label{three_data_matome}
\end{figure*}

Figure~\ref{three_data_matome}~(a) shows the infinite-time average of the probability of the Fourier walk on the three-community network in Fig.~\ref{three_matome}~(a), computed according to Eq.~\eqref{infinitetimeave} based on the numerical diagonalization of $U$. The vertical axis shows the initial node $i$, the horizontal axis shows the target node $l$, and each square color-codes the amplitude of the time-averaged probability $\overline{p(i\to l)}$; note that the probabilities are roughly proportional to the number of links, and those of the hubs (the nodes 1, 13, and 21) are the highest. We can thus identify hubs clearly from the infinite-time average of the probability.

Based on the observation in Fig.~\ref{three_data_matome}~(a), we define the \textit{normalized} probability $P(i\to l;t)$ of each node by dividing the probability $p(i\to l;t)$ by the number of links of the target node $l$:
\begin{align}
P(i\to l;t)&=\frac{p(i\to l;t)}{k_l}\nonumber \\
&=\frac{1}{k_l k_i}\sum_{m=1}^{k_l}\sum_{j=1}^{k_i}\abs{\bra{l\to m}U^t\ket{i\to j}}^2.
\label{normalizedprob-t}
\end{align}
The infinite-time average of the normalized probability is given by
\begin{align}
\overline{P(i\to l)}&=\frac{\overline{p(i\to l)}}{k_l}\nonumber \\
&=\lim_{T\to\infty}\frac{1}{T}\frac{1}{k_l k_i}\sum_{t=0}^{T-1}\sum_{m=1}^{k_l}\sum_{j=1}^{k_i}\abs{\bra{l\to m}U^t\ket{i\to j}}^2
\\
&=\frac{1}{k_l k_i}\sum_{\mu=1}^{D}\sum_{m=1}^{k_l}\sum_{j=1}^{k_i}\abs{\bra{l\to m}\ket{\mu}}^2\abs{\bra{\mu}\ket{i\to j}}^2.
\label{normalizedprob}
\end{align}
This infinite-time average then becomes symmetric with respect to the exchange of $l$ and $i$ as in $\overline{P(i\to l)}=\overline{P(l\to i)}$.

Figure~\ref{three_data_matome}~(b) color-codes the infinite-time average of the \textit{normalized} probability $\overline{P(i\to l)}$ calculated from the eigenvectors of the Fourier walk for the three-community network in Fig.~\ref{three_matome}~(a). The normalized transition probability between the initial node and the other nodes in the same community is high, which reveals the community structure. 

Figure~\ref{three_data_matome}~(c)--(e) shows the same quantity as in Fig.~\ref{three_data_matome}~(b), but only for the cases in which the walk starts from the hubs (the nodes $i=1$, $13$, and $21$, which are indicated by red arrows in Fig.~\ref{three_data_matome}~(c)--(e)). In order to detect the community structure quantitatively, we here tentatively define the threshold for the detection of a community to be $q=1/D$, where $D=78$, which is indeed the stationary probability normalized by the number of links $k_l$ of the {\it classical} random walk on the network. 

We thereby define a community as follows: 
\begin{description}
\item[(i)] We first define a hub $i$ as a node with the largest order $k_i$;
\item[(ii)] If the normalized probability starting from a hub $i$ to a node $l$ is greater than the threshold $q$, namely if
\begin{align}
\overline{P(i\to l)}>q,
\end{align}
the node $l$ is a member of the community of the hub $i$.
\end{description}
This algorithm clearly reveals the three communities of the three-community network in Fig.~\ref{three_matome}~(a).
%; the nodes whose probabilities are higher than the threshold belong to the community whose hub is the initial node. 
For instance, if the Fourier walk starts from the hub 1 as in Fig.~\ref{three_data_matome}~(b), the probability of the nodes 2, 3, 4, 5, 6, 7, which belong to the same community, is higher than the threshold $q$. We can thus successfully identify which community each node belongs to. 

In order to justify the algorithm from a different perspective, we show that the Fourier walk on the network is localized in a community to which the initial node belongs.
Let us evaluate the localization of the eigenvectors using the inverse participation ratio (IPR)~\cite{IPR1, IPR2}. The IPR of an eigenvector
\begin{align}
\ket{\mu}=\sum_{l=1}^N\sum_{m=1}^{k_l}\psi_{\mu}(l,m)\ket{l\to m}
\end{align}
is given by
\begin{align}
\textrm{IPR}(\mu)=\frac{\sum_{l=1}^{N}{p_\mu(l)}^2}{\left(\sum_{l=1}^{N}p_\mu(l)\right)^2}=\sum_{l=1}^{N}{p_\mu(l)}^2,
\end{align}
where the probability $p_\mu(l)$ is
\begin{align}
p_\mu(l)=\sum_{m=1}^{k_l}|\psi_\mu(l, m)|^2
\end{align}
with the normalization $\sum_{l=1}^{N}p_\mu(l)=1$ for all $\mu$.

If the eigenvector is sharply localized to one node, the IPR is close to unity. If the eigenvector is delocalized, the IPR is as small as $1/N$, which is $1/21\approx0.0476$ in the present case of the three-community network in Fig.~\ref{three_matome}~(a). If the eigenvector were localized uniformly in one of the communities of the network as in 
\begin{align}
p_{\mu}(l)=\begin{cases}\dfrac{1}{7}~~~\textrm{for}~~l=1,2,\cdots,7,\\0~~~\textrm{otherwise},\end{cases}
\end{align}
the IPR would be exactly $1/7\approx0.14$.

\begin{figure*}
\centering
\includegraphics[width=0.75\textwidth]{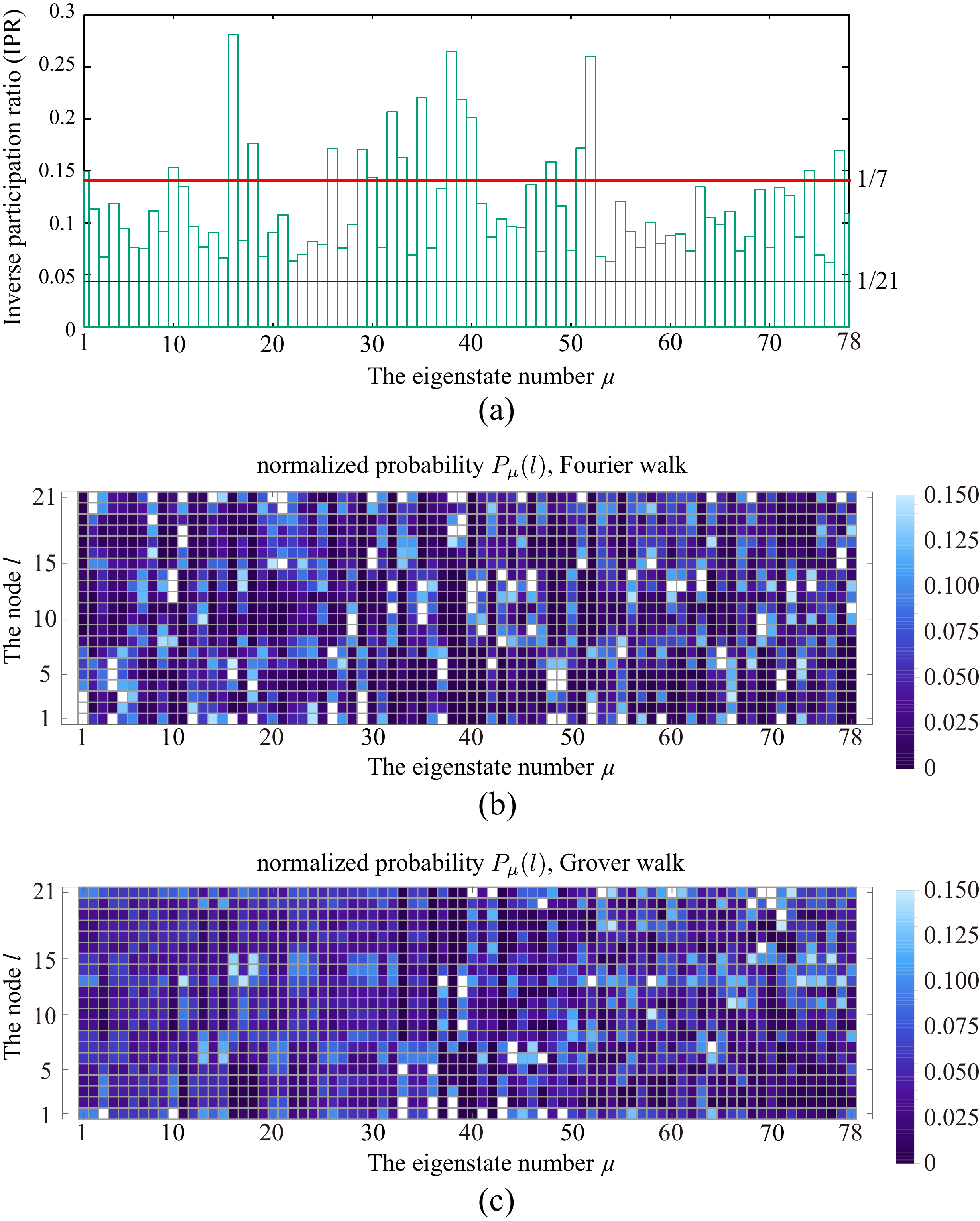}
\caption{(a) The inverse participation ratio (IPR) of each eigenvector for the Fourier walk on the three-community network. The horizontal axis shows the eigenstate number $\mu$ and the vertical axis shows the IPR of each eigenstate. The thin (blue) horizontal line indicates the IPR value $1/21$, which would correspond to a uniformly extended state, while the thick (red) horizontal line indicates the IPR value $1/7$, which would correspond to a state uniformly localized in one of the three communities. (b) The \textit{normalized} probability $P_{\mu}(l)$ in Eq.~\eqref{pmul} calculated from the eigenstates for the Fourier walk on the three-community network. The horizontal axis shows the eigenstate number $\mu$ and the vertical axis shows the node $l$, while each square indicates the value of $P_{\mu}(l)$. (c) The same but for the Grover walk.}
\label{IPR_matome}
\end{figure*}

Figure~\ref{IPR_matome}~(a) shows the IPR of each eigenvector for the Fourier walk on the three-community network. We find that all states have the IPR higher than $1/21\simeq0.04762$ (the thinner horizontal line in Fig.~\ref{IPR_matome}~(a)) and several eigenvectors are localized more strongly than the IPR $=1/7\simeq 0.1429$ (the thicker horizontal line in Fig.~\ref{IPR_matome}~(a)).
The naive average of the IPR over all eigenstates is about $0.1151\simeq 1/8.7$, which is not far from $1/7$.

Figure~\ref{IPR_matome}~(b), on the other hand, shows the \textit{normalized} probability
\begin{align}\label{pmul}
P_{\mu}(l)=\dfrac{p_{\mu}(l)}{k_l}
\end{align}
of the Fourier walk on the three-community network.
The probability distribution for each eigenvalue shows the localization, often over a community. For the probability distribution for the eigenstate number 1 (for which IPR is about 0.150), for instance, the probability for the node 1 and the nodes in the same community (from $l=1$ to $l=7$) is visibly higher than that of the other nodes; in other words, this eigenvector is localized in the first community. Similarly, the probability distribution of the eigenstate number 46 (for which IPR is bout 0.137) is localized in the second community, and that of the eigenstate number 11 (for which IPR is bout 0.135) is localized in the third one.

The localization of the quantum walk may be related to the Anderson localization. In the standard sense, the Anderson localization is the property of quantum particles in random media~\cite{Anderson, randompotential}. There are several studies on the Anderson localization of the discrete-time quantum walk on lattices with randomness~\cite{Andersondiscrete, Andersononedimensional}. The quantum walk on the complex network may be  similar to the quantum particle in random media because of the inhomogeneity of the network, and hence may experience the Anderson localization.

\subsection{Finite-time calculation}
\label{sec3B}
We next present our finite-time results of the quantum walk on the same three-community network in Fig.~\ref{three_matome}~(a). We operated the unitary matrix $U$ to the initial state $\ket{i\to j}$ up to 100 steps and averaged the resulting probability over $j$.

%As can be seen from Fig.~\ref{proave418threestep}, the Fourier walk behaves roughly periodically but not completely. 
Figure~\ref{three_data_matome}~(f) shows the time average of the normalized probability $\overline{P_{t}(i\to l)}$ over 100 steps from $t=1$ through $t=100$. It is almost the same as Fig.~\ref{three_data_matome}~(b), also revealing the community structure. 
The fact that the finite-time average is almost equal to the infinite-time average is presumably thanks to the property of the quantum walk that the front of the probability spreads linearly.
This implies that we can apply the present method to complex networks which are too large to diagonalize the time-evolution unitary matrix $U$ by computing a finite-time average instead of the infinite-time average.

%Let us now compare the numerical results for the Fourier walk and the Grover walk. %Figure~\ref{proave714groverthreestep}~(a)--(l) shows the normalized probability $P(l\to i;t)$ on several steps of the Grover walk on the same network. 
Figure~\ref{three_data_matome}~(g) shows, on the other hand, the same time average of the normalized probability $\overline{P(i\to l;t)}$ but for the \textit{Grover} walk.
% over 100 steps from $t=1$ through $t=100$. 
We can see that the diagonal elements are much larger than the other elements.
In other words, the Grover walk mostly stays at the initial node through the 100 steps, implying strong localization at each node.
%the normalized probability on the initial node is high, but that on the other nodes of the same community is relatively low. 

We may relate this phenomenon to findings for the Grover walk on regular lattices~\cite{Inui04,Segawa14,Komatsu18}.
As we mentioned in Sec.~\ref{sec3A}, it has been proven that the eigenvectors with the degenerate eigenvalues $\pm 1$ of the unitary matrix of the Grover walk on regular lattices are broken down to states localized on loops, which leads to a proof of the localization of the walk on the initial node.
This may be also the case in the present three-community network. % (although it is not trivial to show the localization of each eigenvector because the numerical result generally gives only a superposition of all localized states with the same eigenvalue).
Indeed, the localization numerically demonstrated in Fig.~1~(a) of Ref.~\cite{Inui04} resembles the behavior of the diagonal concentration in Fig.~\ref{three_data_matome}~(g).
Figure~\ref{IPR_matome}~(c) also shows that the normalized probability $P_{\mu}(l)$ in Eq.~\eqref{pmul} calculated from the eigenstates for the Grover walk are localized in nodes rather than in communities if the eigenvalue is $\pm 1$ (the eigenstate numbers from 41 to 78).
For examples, the states 41, 44, 47, 49 and so on, have small elements in a community but are mostly localized to one or a couple of nodes.

\begin{figure}
\includegraphics[width=\columnwidth]{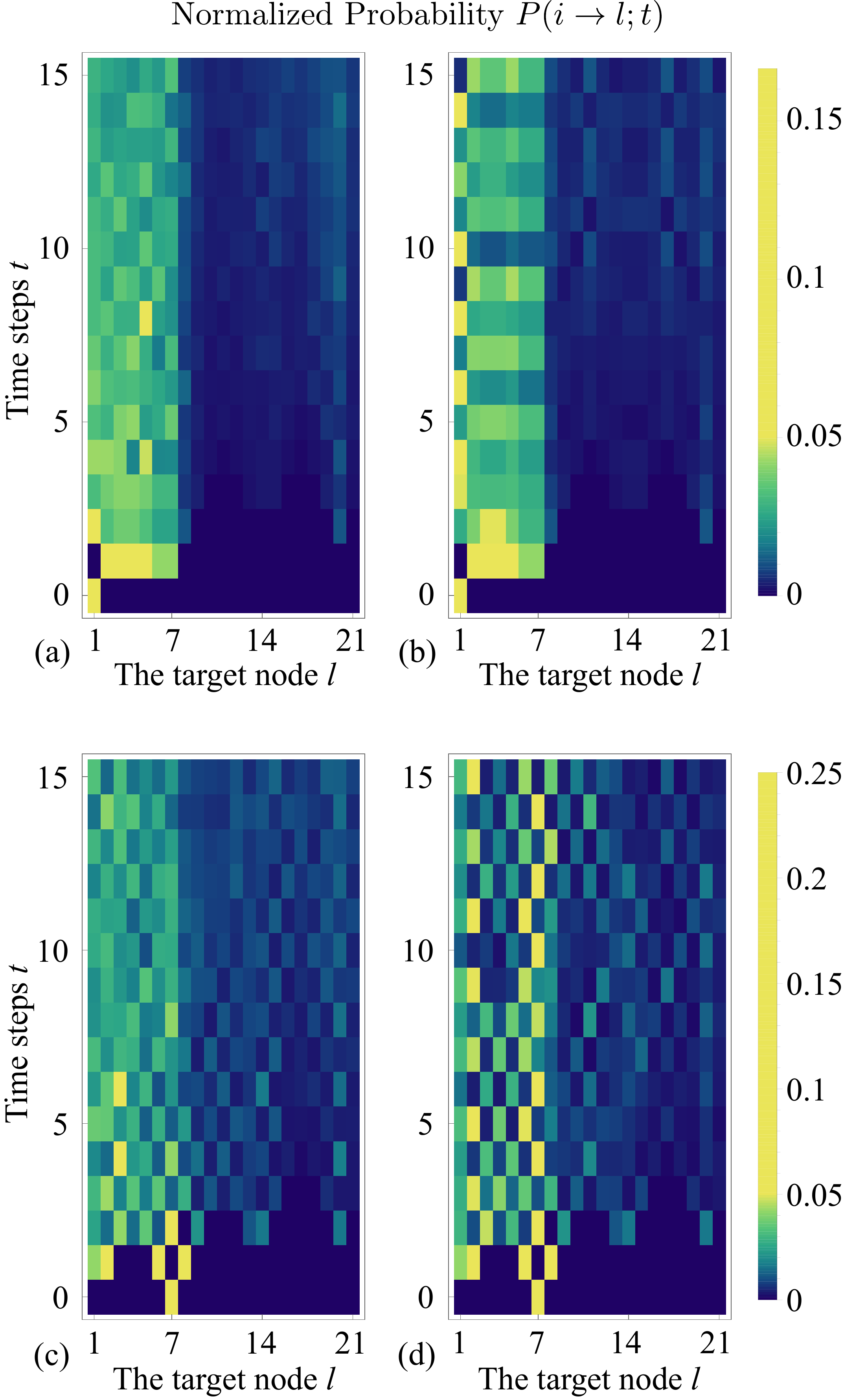}
\caption{The time evolution on the three-community network of the normalized probability $P(i\to l;t)$ from $t=0$ to $t=15$ for (a) the Fourier walk with $i=1$;
(b) the Grover walk with $i=1$;
(c) the Fourier walk with $i=7$;
(d) the Grover walk with $i=7$.}
\label{fig5-1}
\end{figure}

Figure~\ref{fig5-1} shows the first few steps in the time evolution $P(i\to l;t)$ of the Fourier and Grover walks that started from various initial nodes $i$.
When a walk starts from the node 1 of the three-community network in Fig.~\ref{three_matome}~(a), the Fourier walk (Fig.~\ref{fig5-1}~(a)) spreads over the first community in a couple of steps and stays so afterwards.
On the other hand, the Grover walk (Fig.~\ref{fig5-1}~(b)), although it has higher probabilities over the first community than the rest, shows some oscillation in time and has even higher probability at the initial node 1 from time to time.
When a walk starts from the node 7, the Fourier walk (Fig.~\ref{fig5-1}~(c)) again spreads over the first community in afew steps and stays so afterwards. 
The Grover walk (Fig.~\ref{fig5-1}~(d)), however, has high probabilities on nodes 7, 6, and 2.

To summarize in short, the time evolution of the Fourier walk tends to get localized over a community whichever node it starts from, while that of the Grover walk tends to get localized on a couple of nodes around the initial one.
Therefore, the Fourier walk reveals the community structure more clearly than the Grover walk.

%Finally, we show in Fig.~\ref{FP420threestep} the normalized probability $P(l\to i;t)$ on several steps of the classical random walk on the same network. 
Finally, we compare the probability of the quantum walk to that of the classical random walk on the same network.
The probability of the classical random walk eventually relaxes to the flat distribution, which is equal for all nodes, and hence the infinite-time average of the probability does not reveal the community structure. For community detection we would have to choose a specific time step, which is unknown \textit{a priori}. We thus claim that using the time average of the probability of the quantum walk is a more tractable way of community detection than trying to find a specific time step of the classical random walk.

\section{Application to real-world networks}
\label{sec4}
\subsection{Zachary's karate-club network}
\label{sec4A}
\begin{figure*}
\begin{center}
\includegraphics[width=0.85\textwidth]{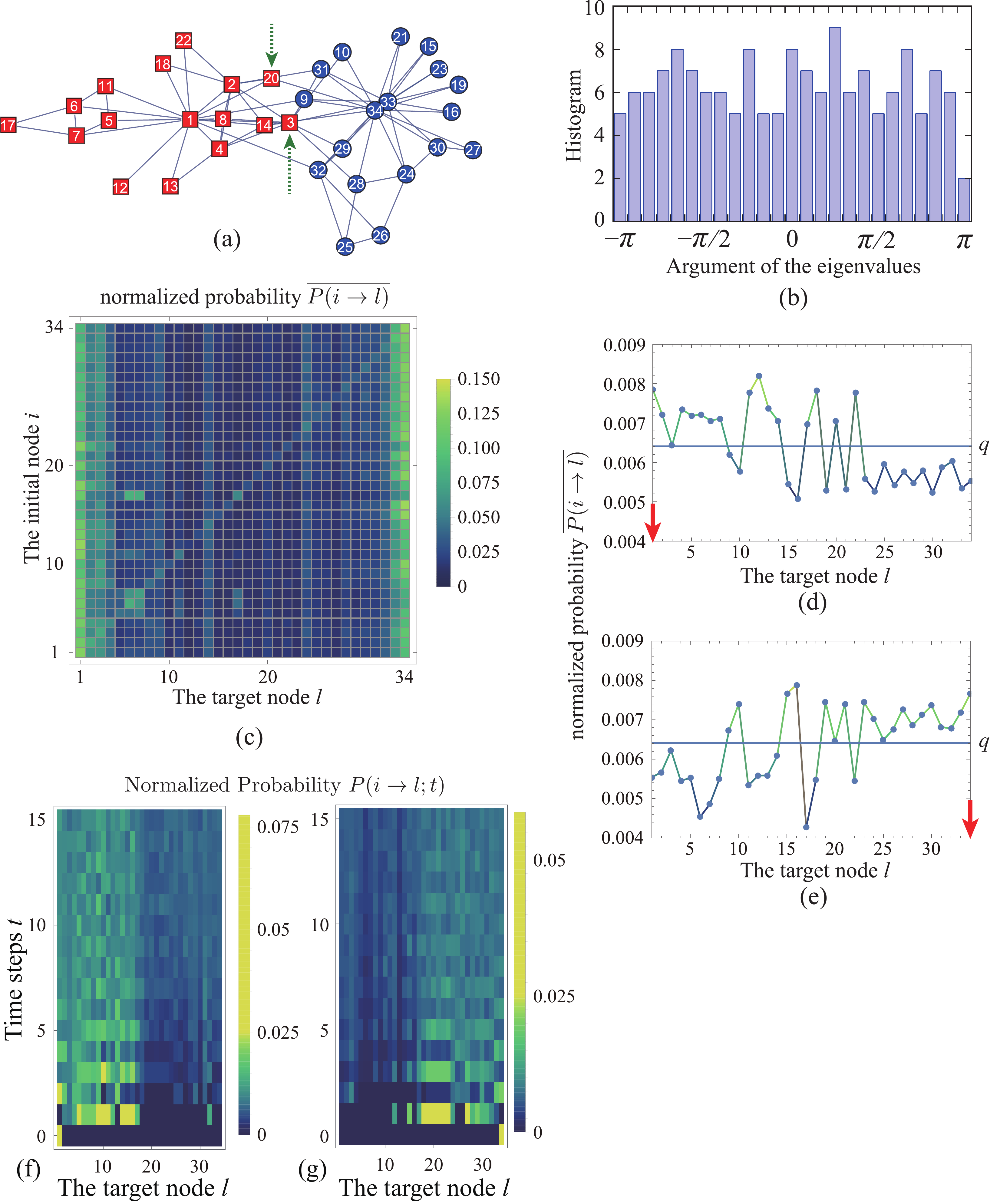}
\end{center} 
\caption{(a) Zachary's karate-club network~\cite{zachary}, for which $N=34$ and $D=156$. The hubs are the nodes 1 and 34. The red square nodes are supposed to be in the group of the node 1 and the blue circular nodes are the group of the node 34. (b) The distribution of the argument of the eigenvalues of the time-evolution unitary matrix of the Fourier walk on the karate-club network. (c) The infinite-time average of the probability $\overline{p(i\to l)}$ in Eq.~\eqref{infinitetimeave} of the Fourier walk on the karate-club network. The vertical axis shows the initial node $i$ and the horizontal axis shows the target node $l$, while each square indicates the value $\overline{P(i\to l)}$. (d)--(e) The infinite-time average of the \textit{normalized} probability $\overline{P(i\to l)}$ in Eq.~\eqref{normalizedprob} of the Fourier walk on the karate-club network that starts from (d) the hub $i=1$ and (e) the hub $i=34$, each of which is indicated by a red arrow. The horizontal axis shows the target node $l$ and the vertical axis shows the normalized probability $\overline{p(i\to l)}$ with $i=1$, $34$. The horizontal line in the middle indicates the threshold $q=1/156 \simeq 0.00641$.
(f)--(g) The time evolution of the normalized probability $P(i\to l;t)$ from $t=0$ to $t=15$ for the Fourier walk with (f) $i=1$ and (g) $i=34$, where we reordered the target nodes $l$ so that the nodes of the first community in (a) may be gathered to the left and those of the second community to the right.}
\label{karate_matome}
\end{figure*}

Let us apply the above algorithm of the community detection to Zachary's karate-club network~\cite{zachary} (Fig.~\ref{karate_matome}~(a)),
which is a friendship network in a karate club in a university in the USA.
The club split into two communities, one clustered around the instructor (node 1) and the other around the administrator (node 34).
In Zachary's psychological experiment, each member of the club answered his/her friends' names and the community to which he/she belongs. 

The network in Fig.~\ref{karate_matome}~(a), for which the total number of nodes is $N=34$ and the total number of links is $D=156$, is based on the first set of answers. The hubs of this network are the nodes 1 and 34. The second set of answers tells us that the communities are as follows:\\
Group of the node 1 : 1, 2, 3, 4, 5, 6, 7, 8, 11, 12, 13, 14, 17, 18, 20, 22 (red squares in Fig.~\ref{karate_matome}~(a)).\\
Group of the node 34 : 9, 10, 15, 16, 19, 21, 23, 24, 25, 26, 27, 28, 29, 30, 31, 32, 33, 34 (blue circles in Fig.~\ref{karate_matome}~(a)).\\
In this sense, this is a rare case of the complex network for which the `correct' answer of the community detection is known,
although the correctness can be disputed; see the last paragraph of the present section.
We will show that our method `correctly' identifies the two communities.

Figure~\ref{karate_matome}~(b) indicates that the eigenvalues for the Fourier walk, which we computed by the numerical diagonalization of the $156\times 156$ matrix, distributes quite evenly on the unit circle.
We confirmed that there is no degeneracy within the numerical double precision.
This guarantees the computation of the infinite-time average given in Sec.~\ref{sec3A} to be valid for the karate-club network too.
Figure~\ref{karate_matome}~(c) shows the infinite-time average of the probability in Eq.~\eqref{infinitetimeave}, which we computed from the numerical diagonalization. The probabilities of the nodes 1 and 34 are higher than any other nodes. We can clearly identify the nodes 1 and 34 as the hubs in this figure. 

Figure~\ref{karate_matome}~(d)--(e) shows the infinite-time average of the \textit{normalized} probabilities~\eqref{normalizedprob} of the Fourier walk which starts from the hubs (the nodes 1 and 34, which are indicated by red arrows in Fig.~\ref{karate_matome}~(d)--(e)). Let us again tentatively define the threshold to be $q=1/D$, where $D=156$. The nodes whose probabilities are higher than the threshold $q$ belong to the community in which the initial node is the hub. For instance, if the Fourier walk starts from the hub 1, the probability of the node 2, which belongs to the same community, is higher than the threshold $q$. We can thus detect which community each node belongs to.

Figure~\ref{karate_matome}~(f)--(g) shows the time evolution of the normalized probabilities~\eqref{normalizedprob-t} of the Fourier walk that starts from the nodes 1 and 34. Here the nodes in the first community are gathered to the left and those in the second one are to the right. We can clearly see that the walk spreads over the respective community in the first couple of steps in the time evolution.

Comments are in order here;
the detection of the node 3 and 20 (highlighted by the dotted arrow in Fig.~\ref{karate_matome}~(a)) are quite marginal.
First, for the node 20, the normalized probability $\overline{P(1\to20)}$ and $\overline{P(34\to20)}$ are both greater than the threshold $q=1/D$.
Nonetheless, the former $\overline{P(1\to20)}\simeq0.007062$ is markedly greater than the threshold $q\simeq0.006410$, while the latter $\overline{P(34\to20)}\simeq0.006451$ is only marginally greater. A slight increase of the threshold $q$ would exclude the possibility of classifying the node 20 to the community of the hub 34.
We thereby conclude that the node 20 should belong to the community of the node 1.
Second, for the node 3, the normalized probability $\overline{P(1\to3)}$ is only slightly greater than the threshold, although for the node 3, $\overline{P(34\to3)}$ is less than the threshold.

There are indeed several views of the grouping for Zachary's network. One research~\cite{karatenewgroup} divided the nodes into three groups; the first is a group of the node 1, the second is a group of the node 34, and the third is a neutral group of the nodes 9, 10, 20, 28, 29. It is therefore reasonable that the node 20 has a marginal value in Fig.~\ref{karate_matome}~(d). In another research~\cite{senseikara}, their algorithm classified the node 3 into the group of the node 34. This is consistent with our result that the node 3 has a marginal value in Fig.~\ref{karate_matome}~(c).
After all, the grouping according to the second set of answers of Zachary's experiment is based on personal views of each subject, and hence is not the only possible answer but remains to be a quite possible one.

\subsection{USA airport network}
\label{sec4B}
\begin{figure*}
\centering
\includegraphics[width=0.8\textwidth]{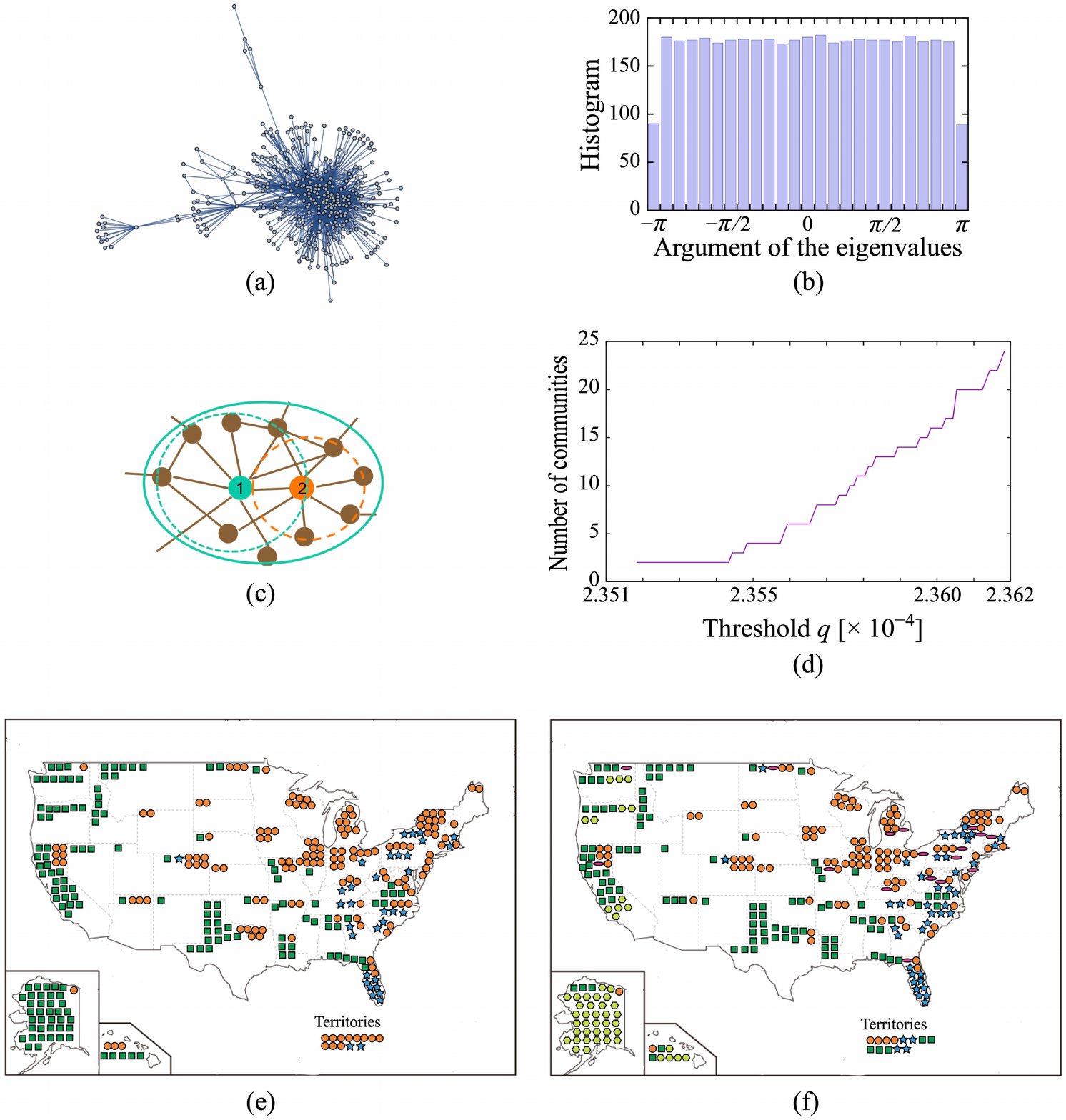}
\caption{(a) The airport transport network in the USA in 1997, for which $N=332$ and $D=4252$. (b) The distribution of the argument of the eigenvalues of the time-evolution unitary matrix for the Fourier walk on the US Airport network. (c) A schematic illustration of the community which has two hubs. The node 1 is the first hub and the node 2 is classified as a member of the group (green broken circle). The node 2 is the second hub concurrently. We classify the hub 2 and its community (orange broken circle) to the community of the hub 1, ending up with a larger community (green solid circle). (d) The number of communities depending on the threshold $q$. The horizontal axis shows the threshold $q$ that we set and the vertical axis shows the number of communities that we obtain.  (e) The result of the community detection of the airport network. We classify the nodes into three communities when we use the threshold $q\simeq0.0002354734$. The orange circle shows the airport which belongs to the first community (147 nodes). The green square shows the second one  (151 nodes) and the blue star shows the third one (34 nodes). (f) The result of the community detection of the airport network. We have five communities when we use the threshold $q\simeq0.0002355834$. The olive hexagon shows the airport which belongs to the fourth community and the red ellipse the fifth. %(g) The dominating airlines at the ten major departure airports that belong to the third community of the result (e).
}
\label{USA_matome}
\end{figure*}

We next apply our method to the domestic airport network in the USA in 1997~\cite{USAdata, naomichi2} (Fig.~\ref{USA_matome}~(a)).
The original data is a weighted network~\cite{USAdata}, but we use the network data as a non-weighted network. Each node of the airport network corresponds to an airport in the USA. They are connected by a link if there is a flight connection between the two airports. The total number of nodes of the airport network is $N=332$ and the total number of links is $D=4252$. 
We computed the infinite-time average~\eqref{infinitetimeave} of the probability of the Fourier walk by the numerical diagonalization of the $4252\times 4252$ matrix. 
Figure~\ref{USA_matome}~(b) shows that the eigenvalues of the time-evolution unitary matrix are distributed almost evenly on the unit circle of the complex plane.
This validates the usage of Eq.~\eqref{Kronecker}.

The community structure of this airport network is \textit{a priori} unknown unlike the prototypical three-community network and Zachary's karate-club network. Based on the successful results above, we here use the following algorithm for community detection:
\begin{description}
\item[(i)]We order the nodes according to the number of links $k_i$, and regard the nodes from the top of the list as candidates for hubs.
\item[(ii)]Starting from the node $i$ with the highest degree, which is the first candidate for the hub, we classify the nodes $l$ whose normalized probability~\eqref{normalizedprob} is higher than a threshold $q$, as in $\overline{P(i\to l)}>q$, into the community of the hub $i$.
\item[(iii)]We carry out (2) repeatedly, ignoring the nodes that have been classified, until all of the nodes are classified into communities. If the node with the highest degree at the moment (\textit{e.g.~}the node 2 in Fig.~\ref{USA_matome}~(c)) has been already classified into a community (\textit{e.g.~}the green broken circle in Fig.~\ref{USA_matome}~(c)), we assume that the hub and the members of the group of the hub (\textit{e.g.~}the orange broken circle in Fig.~\ref{USA_matome}~(c)), belong to the community into which the hub has been classified (\textit{e.g.~}the solid circle in Fig.~\ref{USA_matome}~(c)). 
\end{description}

When we use the threshold $q=1/D\simeq0.0002351834$ as was in the two cases above, we classify all the nodes into two communities, one with 260 nodes, the other with 72. As we can see in Table~\ref{tab-airline}~(a), most of the major airports are classified into the first community, while the second community contains mostly minor airports with a few exceptions. 
\begin{table*}
%\centering
\caption{The top airports in each community according to our algorithm, along with the carrier that carried the largest number of passengers  out of it.
(a) For the threshold $q=1/D\simeq 0.0002351834$, all airports are classified into two communities. We omitted some minor airports in the second community. (b) For the threshold $0.0002354434 \leq q \leq 0.0002354734$, all airports are classified into three communities.}
\label{tab-airline}
\vspace{\baselineskip}
(a) $q=1/D\simeq 0.0002351834$\\[0.5\baselineskip]
\begin{tabular}{ll|ll}
\hline
First community & & Second community  & \\
\hline
Departure Airport & Carrier & Departure Airport & Carrier \\
\hline\hline
Chicago O'Hare, IL (ORD)        & United     & Dallas/Fort Worth, TX (DFW)   & American  \\
Atlanta, GA (ATL)               & Delta      & San Francisco, CA (SFO)       & United    \\
St.\ Louis, MO (STL)            & TWA        & Salt Lake City, UT (SLC)      & Delta     \\
Pittsburgh, PA (PIT)            & US Airways  & Nashville, TN (BNA)           & Southwest \\
Charlotte, NC (CLT)             & US Airways & New York JFK, NJ (JFK)        & American       \\
Denver, CO (DEN)                & United     & Portland, OR (PDX)            & Alaska    \\
\cline{3-4}
Minneapolis-St.\ Paul, MN (MSP) & Northwest  & & \\ %Chicago Midway, IL (MDW)      & Southwest \\
Detroit, MI (DTW)               & Northwest  & & \\ % Long Island
New York Newark, NJ (EWR)       & Continental     & & \\
Philadelphia,PA (PHL)           & US Airways & & \\
Houston, TX (IAH)               & Continental     & & \\
Cincinnati, OH (CVG)            & Delta      & & \\
Phoenix, AZ (PHX)               & America West  & & \\
Los Angeles, CA (LAX)           & United     & & \\
Seattle-Tacoma, WA (SEA)        & Alaska     & & \\
Orlando, FL (MCO)               & Delta & & \\
Baltimore, MD (BWI)             & US Airways & & \\
New York La Guardia, NJ (LGA)   & Delta & & \\
Raleigh, NC (RDU)               & US Airways & & \\
Boston, MA (BOS)                & Delta      & & \\
Las Vegas, NV (LAS)             & Southwest  & & \\
Washington Dulles, VA (IAD)     & United     & & \\
Miami, FL (MIA)                 & American   & & \\
Cleveland, OH (CLE)             & Continental  & & \\
Memphis, TN (MEM)               & Northwest  & & \\
Tampa, FL (TPA)                 & Delta      & & \\
Washington National, VA (DCA)   & US Airways & & \\
Indianapolis, IN (IND)          & US Airways & & \\
\cline{1-2}
\end{tabular}
\\
\vspace{\baselineskip}
(b) $0.0002354434 \leq q \leq 0.0002354734$\\[0.5\baselineskip]
\begin{tabular}{ll|ll|ll}
\hline
First community & & Second community & & Third community & \\
\hline
Departure Airport & Airliner & Departure Airport & Airliner & Departure Airport & Airliner \\
\hline\hline
Chicago O'Hare, IL (ORD)        & United     & Dallas/Fort Worth, TX (DFW)   & American   & Atlanta, GA (ATL)             & Delta      \\
St.\ Louis, MO (STL)            & TWA        & Charlotte, NC (CLT)           & US Airways & Philadelphia,PA (PHL)         & US Airways \\
Pittsburgh, PA (PIT)            & US Airways  & Denver, CO (DEN)              & United     & Cincinnati, OH (CVG)          & Delta      \\
Minneapolis-St.\ Paul, MN (MSP) & Northwest  & San Francisco, CA (SFO)       & United     & Orlando, FL (MCO)             & Delta \\
Detroit, MI (DTW)               & Northwest  & Houston, TX (IAH)             & Continental     & Baltimore, MD (BWI)           & US Airways \\
New York Newark, NJ (EWR)       & Continental     & Los Angeles, CA (LAX)         & United     & Raleigh, NC (RDU)             & US Airways  \\
Phoenix, AZ (PHX)               & America West  & Salt Lake City, UT (SLC)      & Delta      & New York La Guardia, NJ (LGA) & Delta \\
Boston, MA (BOS)                & Delta       & Seattle-Tacoma, WA (SEA)      & Alaska     & Miami, FL (MIA)               & American   \\
Washington Dulles, VA (IAD)     & United     & Nashville, TN (BNA)           & Southwest  & New York JFK, NJ (JFK)        & American     \\
Cleveland, OH (CLE)             & Continental  & Las Vegas, NV (LAS)           & Southwest  & Memphis, TN (MEM)             & Northwest \\
Indianapolis, IN (IND)          & US Airways & Washington National, VA (DCA) & US Airways & Tampa, FL (TPA)               & Delta     \\
%                             Portland, OR (PDX)            & Alaska     &
\hline
\end{tabular}
\end{table*}

Changing the value of the threshold $q$ reveals the hierarchical structure of the communities as the dendrogram in Fig.~\ref{dendrogramcommunity} implies; see Fig.~\ref{USA_matome}~(d). We find three communities when we use the threshold $q$ in the range $0.0002354434\leq q\leq0.0002354734$ (see Fig.~\ref{USA_matome}~(e)). The orange circle shows the airport which belongs to the first community (147 nodes), the green square the second (151 nodes), and the blue star the third (34 nodes). Comparing the top airports in Table~\ref{tab-airline}~(a) and~(b), we see that many major airports in the first community in Table~\ref{tab-airline}~(a) are distributed to the second and third communities in Table~\ref{tab-airline}~(b).

The top airport of each community is the hub airport of the present-day three major airline companies, Chicago O'Hare for the United, Dallas/Fort Worth for the American, and Atlanta for Delta.
We therefore claim that each of the three communities indicate the subnetwork of airline companies.
Nonetheless, except for the hub airports, we see mixtures of various airlines.
Note that TWA, US Airways, and America West have been merged into the American,
while Northwest to Delta and Continental to the United.
We could observe that the these mergers were strategically reasonable in the sense that Delta and the American respectively merged  companies that appear in communities different from their own hub airports.
It would be interesting to analyze the airport network after mergers (including the one of TWA into the American), but it is out of the scope of the present paper.

%We claim that the three communities correspond to the three major airline companies, namely the United, the American and the Delta. (In the case of finding two communities, the third is merged into the first.) We can see this as follows.
%First, the hub of each community is indeed the hub airport of the respective company; Chicago O'Hare International for the first community, Dallas/Fort Worth International for the second, and Hartsfield-Jackson Atlanta International for the third.
%Second, we can confirm that other airports of each community are indeed dominated by the respective company.
%For example, most of the top ten airports in the third community (see Fig.~\ref{USA_matome}~(g)) are dominated by the Delta Airlines~\cite{airlines}.

Our algorithm of community detection can further find the hierarchy of the airline companies. By increasing the threshold further to $q=0.0002355834$, we find five communities (see Fig.~\ref{USA_matome}~(f)), with 109, 111, 51, 44, 17 nodes, respectively.
The fourth community (olive hexagons) splits off exactly from the second community in Fig.~\ref{USA_matome}~(e), while the fifth community (red ellipses) mostly from the first.
We can easily see that the fourth community corresponds to the Alaska Airlines~\cite{airlines}, which is indeed a partner company of the American Airlines, a major company of the second community in Fig.~\ref{USA_matome}~(e).

In contrast, a previous research~\cite{naomichi2} divided the airports into two communities that geographically corresponds to the east and the west, the latter including the midwest.
Our algorithm excels in finding a different structure since it starts from finding the hubs. 

\section{Conclusion}
\label{sec5}
In the present paper, we defined the discrete-time quantum walk on complex networks and utilized it for community detection. We numerically showed that the Fourier walk is localized in a community to which the initial node belongs. We calculated the infinite-time average of the transition probability by the use of the eigenvectors. We confirmed that the eigenvectors of the Fourier walk tend to be localized in a community, while those of the Grover walk tend to be localized in some specific nodes.

We found that the infinite-time average reveals the community structure better if the eigenvalues of the unitary matrix are non-degenerate, and hence the Fourier walk is more suitable for community detection than the Grover walk. The transition probability becomes higher in proportion to the number of links, and thereby we can detect the hubs. Next, we normalized the probability of each node by dividing it by the number of links. The normalized probability in the initial node and the other nodes in the same community is high, which reveals the community structure. Meanwhile, the probability of the classical random walk on the same network eventually converges to the flat distribution. We thus claim that the time average of the probability of the Fourier walk on complex networks reveals the community structure more explicitly than that of the classical random walk.

Finally, we applied the method to the real-world networks. For Zachary's karate-club network, we confirmed that our method reveals its community structure correctly. Most nodes of the network are classified clearly, while two nodes are marginally identified. This result is consistent with other researches. For the airport network in the USA, we confirmed that our method reveals its community structure that corresponds to the three major airline companies in the USA. By adjusting the threshold, our algorithm successfully reveals the hierarchical structure of the communities as the dendrogram in Fig.~\ref{dendrogramcommunity} implies. 

We argued that the strong localization of the Grover walk is presumably due to many eigenstates of degenerate eigenvalues $\pm 1$, which were mathematically proven to localized on loops~\cite{Segawa14};
hence we are almost certain that the Grover walk is not suitable for community detection.
On the other hand, we numerically showed that the Fourier walk works for community detection, but we are yet to find any mathematical reasons why it does.
We have not tried other types of quantum walks either. 
These are beyond the scope of the present single paper, and should be pursued in future studies.

Let us finally add a remark on a possible extension of the present algorithm. We have defined our quantum walk ignoring the weight and the direction of the links of the networks. We can vary the weight as integers by making each link have multiple connections. In order for a directed network to accommodate a quantum walk, the network cannot have any dead ends of directed links~\cite{directed}. We may be able to apply our algorithm to the directed network as long as the condition is satisfied.

\section*{Acknowledgements}
We would like to show our great appreciation to Prof.~Hideaki Obuse. He taught us about quantum walks in detail, and gave us many pieces of advice for our research. We also would like to show our great appreciation to Prof.~Masaki Sano and Prof.~Takeo Kato. They gave us many comments. We also would like to thank many people for many discussions in our poster presentation.
The present study is partially supported by JSPS Grant-in-Aid for Scientific Research (A) No.~19H00658.

\appendix
\section{Transformation to the standard shift operator}
\label{AppA}

In the present Appendix, we explicitly show for a one-dimensional lattice that we can transform the atypical shift operator~\eqref{shiftdefinition} to the standard shift operator by multiplying the coin operator by the flipping operator.

Following the main text, let us express the right and left movers on the one-dimensional lattice as follows, respectively:
\begin{align}
\mbox{Right mover: } \cdots, &\ket{(x-1)\to x}, 
\nonumber\\
&\ket{x\to (x+1)}, 
\nonumber\\
&\ket{(x+1)\to(x+2)},\cdots,
\\
\mbox{Left mover: } \cdots, &\ket{(x+2)\to(x+1)},
\nonumber\\
&\ket{(x+1) \to x}, 
\nonumber\\
&\ket{x\to(x-1)}. \cdots,
\end{align}
We can express the atypical shift operator~\eqref{shiftdefinition} in the form of the following matrix:
\begin{align}
S=
%\begin{array}{rl}
%& 
%\begin{array}{cc|cc|cc}
%\bra{(x-1)\to(x-2)} & \bra{(x-1)\to x} & \bra{x\to(x-1)} & \bra{x\to(x+1)} & \bra{(x+1)\to x} & \bra{(x+1)\to(x+2)}
%\end{array}
%\\
%\begin{array}{r}
%\ket{(x-1)\to(x-2)} \\ \ket{(x-1)\to x} \\ \hline \ket{x\to(x-1)} \\ \ket{x\to(x+1)} \\  \hline \ket{(x+1)\to x} \\ \ket{(x+1)\to(x+2)}
%\end{array}
%&
\left(\begin{array}{c|cc|cc|cc|c}
\ddots & 1 & & & & & \\
\hline
1 & & & & & & & \\
& & & 1 & & & & \\
\hline
& & 1 & & & & & \\
& & & & & 1 & & \\
\hline
& & & & 1 & & & \\
& & & & & & & 1 \\
\hline
& & & & & & 1 & \ddots
 \end{array}
 \right)
 %\end{array}
 \end{align}
under the following ordering of the bases 
\begin{align}
\left(
\begin{array}{l}
\multicolumn{1}{c}{\vdots} \\
\hline
\ket{(x-1)\to(x-2)} \\ \ket{(x-1)\to x} \\ \hline \ket{x\to(x-1)} \\ \ket{x\to(x+1)} \\  \hline \ket{(x+1)\to x} \\ \ket{(x+1)\to(x+2)} \\
\hline
\multicolumn{1}{c}{\vdots}
\end{array}\right)
\end{align}
because
\begin{align}
S\ket{x\to(x-1)}&=\ket{(x-1)\to x},
\nonumber\\
S\ket{(x-1)\to x}&=\ket{x\to(x-1)},
\nonumber\\
S\ket{(x+1)\to x}&=\ket{x\to (x+1)},
\nonumber\\
S\ket{x\to (x+1)}&=\ket{(x+1)\to x}.
\end{align}
The time-evolution unitary operator is therefore given by 
\begin{align}
U=SC
\end{align}
with a coin operator
\begin{align}
C=\cdots\oplus C_{x-1}\oplus C_x \oplus C_{x+1} \cdots
\end{align}
with a $2\times 2$ unitary matrix $C_x$; for example, a Fourier coin
\begin{align}\label{app-fourier}
C_x=\frac{1}{\sqrt{2}}\begin{pmatrix}
1 & 1 \\
1 &-1
\end{pmatrix}
\end{align}

We now define a new coin operator with an additional factor $P_x$ inserted to the left of the original coin operator $C_x$:
\begin{align}
C'=PC=\cdots\oplus P_{x-1}C_{x-1}\oplus P_xC_x \oplus P_{x+1}C_{x+1} \cdots,
\end{align}
where the new factor
\begin{align}
P_x=\begin{pmatrix}
0 & 1 \\
1 & 0
\end{pmatrix}
\end{align}
flips the direction of the right and left movers.
In the new time-evolution operator
\begin{align}
U'=SC'=SPC,
\end{align}
we find 
\begin{align}
S'=SP=\left(\begin{array}{c|cc|cc|cc|c}
\ddots & & 1& & & & & \\
\hline
& & & & & & & \\
& & & & 1 & & \\
\hline
& 1 & & & & & &\\
& & & & & & 1 &\\
\hline
& & & 1 & & & &\\
& & & & & & & \\
\hline
& & & & & 1& & \ddots
\end{array}
\right),
\end{align}
because
\begin{align}
S'\ket{x\to(x+1)}&=S\ket{x\to(x-1)}=\ket{(x-1)\to x},
\nonumber\\
S'\ket{(x-1)\to (x-2)}&=S\ket{(x-1)\to x}=\ket{x\to(x-1)},
\nonumber\\
S'\ket{(x+1)\to (x+2)}&=S\ket{(x+1)\to x}=\ket{x\to (x+1)},
\nonumber\\
S'\ket{x\to (x-1)}&=S\ket{x\to(x+1)}=\ket{(x+1)\to x}.
\end{align}
We can see that the operator $S'$ works as the standard shift operator, which shifts the right mover to the right keeping it as a right mover and shifts the left mover to the left keeping it as a left mover.

Therefore, the time-evolution operator
\begin{align}
U'=SC'=S'C
\end{align}
is the atypical shift operator~\eqref{shiftdefinition} multiplied by a slightly jammed coin operator, for example, for Eq.~\eqref{app-fourier},
\begin{align}
C'_x=PC_x=\frac{1}{\sqrt{2}}
\begin{pmatrix}
1 & -1 \\
1 & 1
\end{pmatrix},
\end{align}
but at the same time, is the standard shift operator $S'$ multiplied by the standard coin operator $C$.
In this sense, the atypical shift operator~\eqref{shiftdefinition} is quite similar to the standard shift operator.

\section{Eigenvectors of the Grover walk on graphs for the eigenvalues $\pm 1$}
\label{AppB}

We here present tutorial examples of the eigenvectors of the Grover walk on graphs for the degenerate eigenvalues $\pm 1$.
The following is based on private discussions with H.~Obuse~\cite{Obuse} and E.~Segawa~\cite{Segawa}.

Let us first note that the Grover coin $C^{\mathrm{G}}_{i}$ in Eq.~\eqref{GroverCoin} always has an eigenvalue $-1$ for an eigenvector with only two nonzero elements.
We can straightforwardly confirm it by applying the Grover coin to the vector $\left(\begin{array}{ccccc} 1 & -1 & 0 & 0 & \cdots\end{array}\right)^T$:
\begin{align}\label{CG}
\frac{1}{k_i}\mqty(
2-k_i&2&2&2&2\\
2&2-k_i&2&2&2\\
2&2&2-k_i&2&2\\
\vdots&\vdots&\vdots&\ddots&\vdots\\
2&2&2&\cdots&2-k_i
)
\mqty(
1 \\
-1 \\
0 \\
\vdots \\
0
)
=\mqty(
-1 \\
1 \\
0 \\
\vdots \\
0
).
\end{align}

\begin{figure*}
\includegraphics[width=0.8\textwidth]{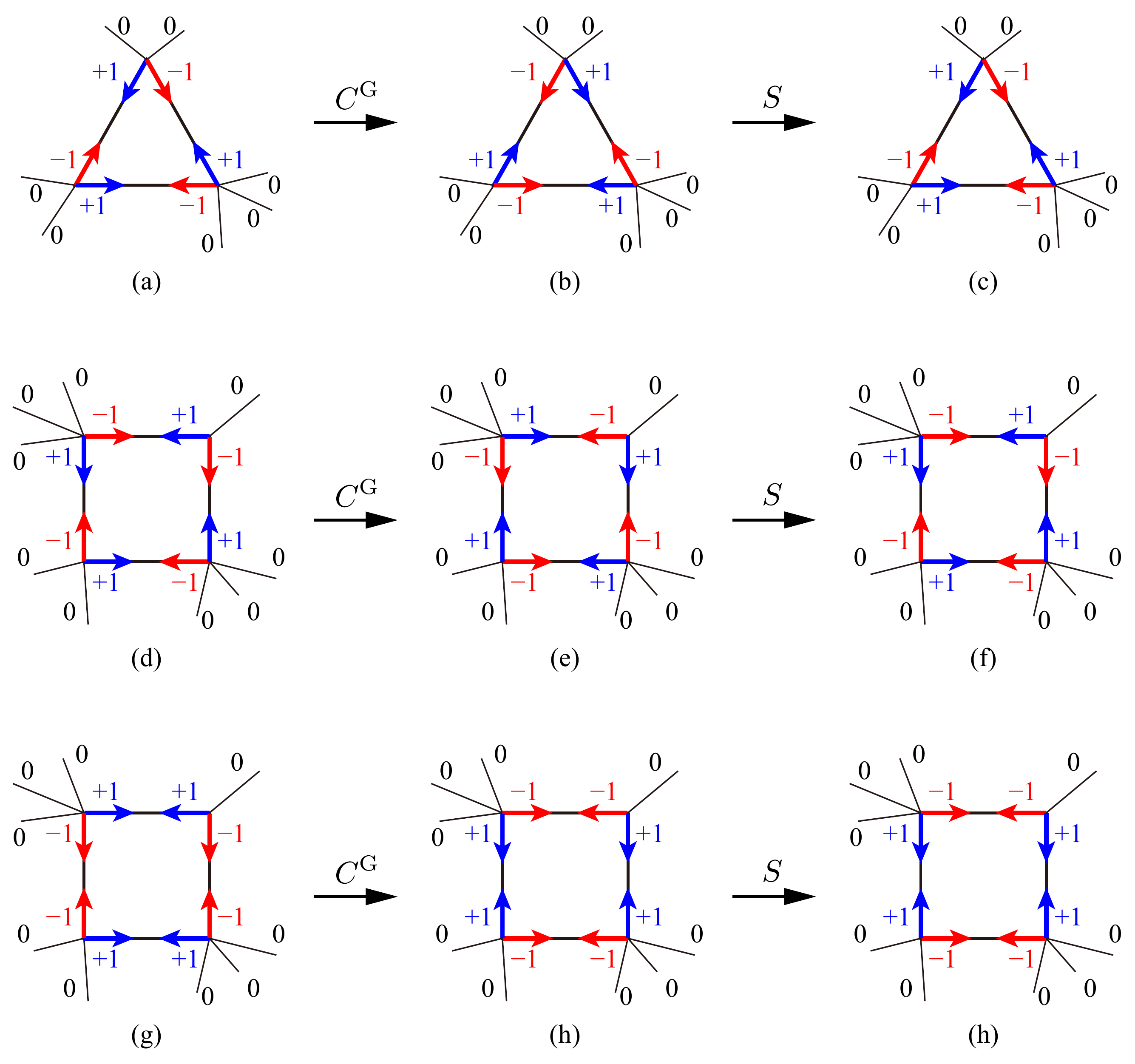}
\caption{(a)-(c) An eigenvector of the Grover walk on a graph, which is localized on a triangular loop, has the eigenvalue $+1$.
(d)-(f) An eigenvector localized on a square loop also has the eigenvalue $+1$.
(g)-(i) Another eigenvector localized on a square loop has the eigenvalue $-1$.}
\label{fig8}
\end{figure*}
We first show~\cite{Obuse} that the vector depicted in Fig.~\ref{fig8}~(a) is an eigenvector of the Grover walk with the eigenvalue $+1$.
Here the (blue) arrow with the sign $+1$ indicates that the vector has an element $+1$ for the basis at the node to which the arrow is attached and being about to hop to the next node. 
The (red) arrow with the sign $-1$ indicates an element $-1$ for the corresponding basis.
The other elements are all zero.
In other words, this vector is strictly localized on a triangular loop.

Application of the Grover coin to the vector changes it to the one depicted in Fig.~\ref{fig8}~(b) because of the operation in Eq.~\eqref{CG}.
Further application of the shift operator in Eq.~\eqref{shiftdefinition} changes it back to the original one as in Fig.~\ref{fig8}~(c).
We have thereby confirmed that the vector in Fig.~\ref{fig8}~(a) is an eigenvector of the Grover walk with the eigenvalue $+1$.

We can confirm in the same way that the vector depicted in Fig.~\ref{fig8}~(d), localized strictly on a square loop is also an eigenvector of the Grover walk with the eigenvalue $+1$.
On a square loop, we can find another vector, depicted in Fig.~\ref{fig8}~(g), is an eigenvector of the Grover walk but with the eigenvalue $-1$.
We can thus easily guess the Grover walk must have large degrees of degeneracies for the eigenvalues $\pm 1$.

Indeed, it has been proven~\cite{Segawa14,Segawa} for the Grover walk on a graph $G$ that the the degeneracy of the eigenvalue $+1$ is $b_1(G)+1$, whereas the degeneracy of the eigenvalue $-1$ is $b_1(G)+1$ if the graph $G$ is bipartite and $b_1(G)-1$ if not, where $b_1(G)=|E|-|V|+1$ is the Betti number of the graph $G$ with $|E|$ and $|V|$ denoting the number of edges (links) and vertices (nodes), respectively.

\begin{figure}
\includegraphics[width=\columnwidth]{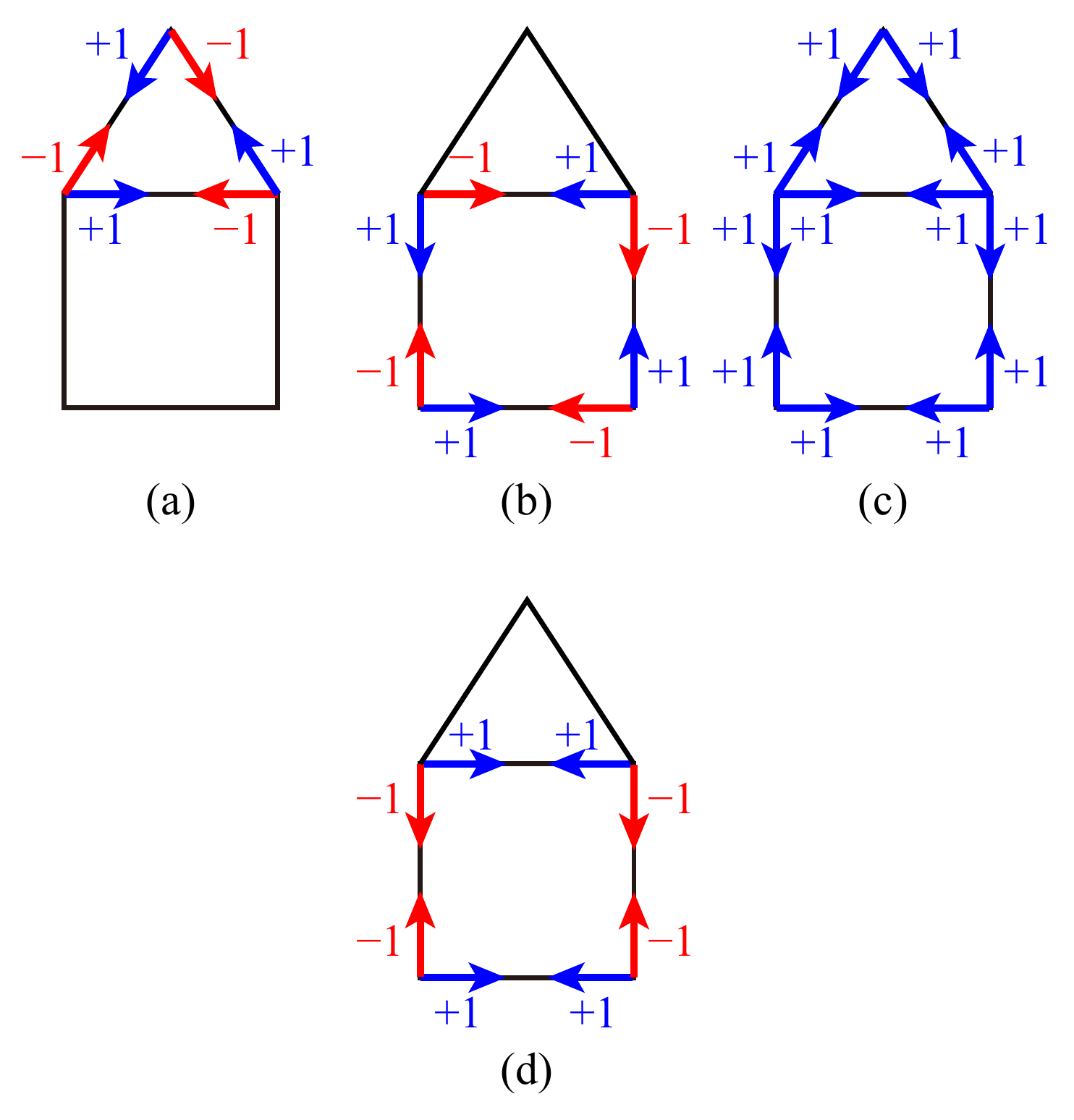}
\caption{(a)-(c) The three eigenvectors degenerate to the eigenvalue $+1$.
(d) The only eigenvector with the eigenvalue $-1$.}
\label{fig9}
\end{figure}
We can easily confirm this \textit{e.g.}\ for the graph in Fig.~\ref{fig9}, which is a combination of a square and a triangle with the Betti number $b_1(G)=6-5+1=2$.
According to the theorem, the degeneracies of the eigenvalues $\pm 1$ are $b_1(G)+1=3$ and $b_1(G)-1=1$, respectively, because the graph is not bipartite.
Indeed, the vectors depicted in Fig.~\ref{fig9}~(a)-(c) have the eigenvalue $+1$, while the vector in Fig.~\ref{fig9}~(d) the eigenvalue $-1$.

Note that although there is always an extended eigenvector, such as exemplified in Fig.~\ref{fig9}~(c), namely, the same element in all bases, with the eigenvalue $+1$, its overlap with the initial state of the Grover walk should be order of $1/\sqrt{D}$ because of the normalization of the eigenvector, and hence we can ignore its contribution for large networks.

For the three-community network in Fig.~\ref{three_matome}~(a), because the Betti number is given by $b_1(G)=39-21+1=19$, the degeneracies in the eigenvalues $\pm 1$ are 20 and 18, respectively. 
For Zhachary's karate club in Fig.~\ref{karate_matome}~(a), they are 46 and 44, and for the airport transport network in Fig.~\ref{USA_matome}~(a), they are 1796 and 1794.
Except for the one extended eigenvector, they are all localized on loops at least in one set of linear combinations of degenerate eigenvectors.

\end{document}